# Photoinduced Magnetic Nanoprobe Excited by Azimuthally Polarized Vector Beam

Caner Guclu, Mehdi Veysi, and Filippo Capolino

*Abstract*—The concept of magnetic nanoprobes (or magnetic nanoantennas) providing a magnetic near-field enhancement and vanishing electric field is presented and investigated, together with their excitation. It is established that a particular type of cylindrical vector beams called azimuthally electric polarized vector beams yield strong longitudinal magnetic field on the beam axis where the electric field is ideally null. These beams with an electric polarization vortex and cylindrical symmetry are important in generating high magnetic to electric field contrast, i.e., large local field admittance, and in allowing selective excitation of magnetic transitions in matter located on the beam axis. We demonstrate that azimuthally polarized vector beam excitation of a photoinduced magnetic nanoprobe made of a magnetically polarizable nano cluster leads to enhanced magnetic near field with resolution beyond diffraction limit. We introduce two figures of merit as magnetic field enhancement and local field admittance normalized to that of a plane wave that are useful to classify magnetic nanoprobes and their excitation. The performance of magnetic nanoprobe and azimuthal polarized beams is quantified in comparison to other illumination options and with several defect scenarios.

## I. INTRODUCTION

NATURAL magnetism at optical frequencies is rather weak when compared to electric response of matter [1]. Indeed optical spectroscopy and microscopy systems mainly work based on electric dipolar transitions in matter rather than their magnetic counterparts. On the other hand, even though natural optical magnetism vanishes, metamaterials with equivalent magnetic dipolar responses have been widely studied in the past decade. For example, arrays of magnetic meta atoms are employed in engineering bulk effective permeability [2]–[7]. Several studies have been devoted to generating artificial magnetism (i.e., effective relative permeability different from unity) for such structures leading to effective permeability engineering. However, artificial magnetism for magnetic near-field enhancement is a rather newer subject of research and its application to boost the weak natural magnetism in matter at the short wavelength range of the electromagnetic spectrum for microscopy applications is rather unexplored yet.

"This work was supported by the Keck Foundation, USA".
All authors are with the Department of Electrical Engineering and Computer Science, University of California, Irvine, CA 92697 USA. (e-mails: cguclu@uci.edu, mveysi@uci.edu, f.capolino@uci.edu).

The building blocks of metamaterials are scatterers that possess scattering modes modeled via multipolar expansion. The magnetic dipolar term in the scattering multipolar expansion of such scatterers is always present alongside the electric dipolar response. Magnetic resonances in such meta atoms have been used also to generate high-quality resonances owing to the reduced radiative losses. In particular, clusters of plasmonic nanoparticles such as the spherical constellations in [8]–[10] have been suggested to generate effective bulk permeability when arranged in array configuration. Also circular clusters of plasmonic particles, which are of interest in this paper, have been suggested with the purpose of engineering negative effective permeability [5], for achieving Fano resonances [11]–[15], and in the quest for detectable photoinduced magnetic forces via artificial magnetism [16]. The characterization of the near-field signature of magnetic nanoprobes may be functional in adding an extra dimension to optical spectroscopy using photoinduced force microscopy [17]. The extra dimension based on magnetic near-field signature could be provided in addition to the electric dipolar near-field signature.

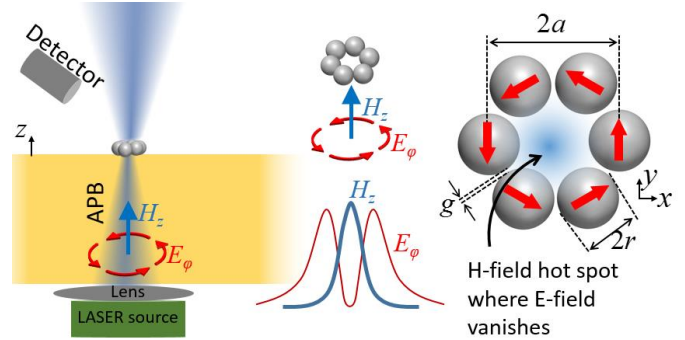

Fig. 1. Illustration of the exemplary setup in which generation of a large magnetic to electric field contrast could be beneficial in detection of (or interaction with) a weakly magnetic response of a matter sample placed at the center of the cluster. The nanoantenna studied here, called magnetic nanoprobe, is made of a resonating circular cluster excited by an APB with longitudinal magnetic field and it generates a strong magnetic field at its center.

The utilization of a magnetic nanoprobe for enhancing the magnetic transitions and suppressing the electric dipolar ones in matter requires: (i) excitation of the magnetic mode of the nanoprobe leading to enhanced magnetic near field, (ii) suppressing the electric near field where magnetic field is enhanced. To help in the latter magnetic to electric field contrast aspect, symmetry in the magnetic nanoprobe and in its excitation plays an important role as we show in this paper. This entails the suppression of the electric dipolar mode and the rest of the higher order electric multipoles in the magnetic nanoprobe. In the quest of selective excitation of the magnetic dipolar mode of the nanoprobe, we turn our attention to vector beams with cylindrical symmetry, so-called cylindrical vector beams [18]–[21], which have been proved to be experimentally functional in the selective excitation of Mie resonances in dense dielectric particles [22]. Cylindrically symmetric vector beams with spatially-dependent electric field vectors, namely radially [18]–[21], [23]–[27] and azimuthally [18]–[21], [28]–[31] polarized vector beams, have been thoroughly investigated specially under tight focusing [23], [26]. A particular cylindrically symmetric vector beam category useful for selective excitation of the magnetic dipolar moment is the azimuthally electric polarized vector beams which hosts a strong longitudinal magnetic field along the beam axis where electric field vanishes [28]–[31]. In the following, we call such beams simply as *azimuthally polarized beam (APB)* referring to the local orientation of their vector electric field. An APB does not only have the capability to selectively excite the magnetic dipolar moment of a magnetic nanoprobe, but also to boost intensity and resolution of the magnetic near-field scattered by a nanoprobe. Owing to the rotational symmetry of the APB and nano probe setup (Fig. 1), one can obtain large magnetic to electric field ratio, denoted also as local field admittance, around the magnetic nanoprobe center when aligned with the beam axis. In [31], the authors quantify electric and magnetic fields of the tightly focused APB within a region with a very large local field admittance and report that scattering by a dense dielectric nanosphere placed in the focal plane of the focused APB leads to enhanced magnetic field with resolution below the diffraction limit.

In this paper, we elaborate on the excitation of a magnetic nanoprobe consisting of a cluster of plasmonic nanospheres which provides significant accessible area with enhanced magnetic field for placing matter samples (i.e., molecules, quantum dots, etc.). The near field of such a plasmonic nano clusters is characterized in terms of newly introduced figures of merit quantifying the magnetic and electric field enhancements and the magnetic to electric field contrast, i.e., the local field admittance. The magnetic nano cluster therefore significantly boosts the total magnetic field and increases spatial magnetic resolution. Such nano clusters when excited with an APB could be useful in boosting the magnetic dipolar transitions of materials located at the cluster's center which are in general weak and overshadowed by stronger electric dipolar transitions. The setup proposed in this paper is depicted in Fig. 1 where a magnetic-dominant region with a strong magnetic field and a vanishing electric field is generated.

## II. LARGE LOCAL FIELD ADMITTANCE AND ENHANCED MAGNETIC FIELD

In the selective excitation of the magnetic dipolar transitions with magnetic-based spectroscopic applications in mind, the main goal is to investigate the physics of magnetic field enhancement within a region where electric field vanishes, so-called magnetic-dominant region. To this purpose we investigate a circular cluster of plasmonic nanospheres that supports a "magnetic" resonance, excited by an APB, whose electric field vanishes on the beam axis aligned with the cluster center. These kinds of nanoantennas are here called magnetic nanoprobes because they are used to enhance the magnetic near field. Several concepts developed in this paper for a circular cluster of plasmonic nanospheres are also applicable to other kinds of magnetic nanoantennas.

We introduce some figures of merit to characterize the quality of magnetic nanoprobes and their excitation. The goal is to quantify the magnetic field enhancement and the magnetic to electric field ratio, i.e., the absolute value of local field admittance normalized by that of a plan wave $1/\eta = \sqrt{\varepsilon/\mu}$, defined as

$$F_H = \frac{\left|\mathbf{H}^{\text{tot}}(\mathbf{r})\right|}{\left|\mathbf{H}^{\text{ext}}(\mathbf{r})\right|}, \quad F_Y = \frac{\eta\left|\mathbf{H}^{\text{tot}}(\mathbf{r})\right|}{\left|\mathbf{E}^{\text{tot}}(\mathbf{r})\right|}, \quad (1)$$

where the superscripts "tot" and "ext" refer to the total field and external excitation field, respectively. For completeness we define also the electric field enhancement as

$$F_E = \frac{\left|\mathbf{E}^{\text{tot}}(\mathbf{r})\right|}{\left|\mathbf{E}^{\text{ext}}(\mathbf{r})\right|}, \quad (2)$$

although in this paper the goal is to boost only the figures of merit in (1). The subscripts of figures of merit *H*, *E*, and *Y* stand for magnetic field, electric field and local field admittance, respectively. In this paper bold fonts denote phasor vector quantities with time-harmonic convention $\exp(-i\omega t)$ where $\omega$ and $t$ refer to real angular frequency and time, respectively. A hat "^" is used to denote unit vectors.

To have large values of figures of merit, we propose to use an APB to illuminate the magnetic nanoprobe as in Fig. 1. The external electric field of the APB is given by the expression

$$\mathbf{E}^{\text{APB}}(\mathbf{r}) = \frac{-i}{\sqrt{2}}\left[\frac{\hat{\mathbf{x}}+i\hat{\mathbf{y}}}{\sqrt{2}}E_{l=-1}^{\text{LG}}(\mathbf{r}) - \frac{\hat{\mathbf{x}}-i\hat{\mathbf{y}}}{\sqrt{2}}E_{l=+1}^{\text{LG}}(\mathbf{r})\right] = \\ = \hat{\boldsymbol{\varphi}}E_{\varphi}^{\text{APB}} = \hat{\boldsymbol{\varphi}}E_{l=\pm 1}^{\text{LG}}e^{\mp i\varphi}, \quad (3)$$

where $E_l^{\text{LG}}$ is the field expression of a Laguerre-Gaussian beam with orbital angular momentum (OAM) order of *l*, and radial mode number *p* = 0 that propagates in the +*z* direction. Note that the choice of + or - sign in (3) is irrelevant. The APB ideally does not possess longitudinal electric field anywhere, while possessing a longitudinal magnetic field that reaches its maximum on the beam axis. On the beam axis, for symmetry reasons, there are no transverse electric and transverse magnetic fields [29]. Note that Laguerre-Gaussian beams are solutions to the wave equation under paraxial approximation [32]. The expression of the Laguerre-Gaussian mode with OAM order $l = \pm 1$ and radial order $p = 0$ is [29]

$$E_{l=\pm 1}^{\text{LG}}(\mathbf{r}) = V \frac{2\rho}{\sqrt{\pi w^2}} e^{-\left(\frac{\rho}{w}\right)^2} \zeta \, e^{-2i\tan^{-1}\left(\frac{z}{z_R}\right)} e^{\pm il\varphi} e^{ikz},$$

$$w = w_0\sqrt{1+\left(\frac{z}{z_R}\right)^2}, \quad \zeta = \left(1 - i\frac{z}{z_R}\right) \quad (4)$$

where $\rho = \sqrt{x^2 + y^2}$ and $\varphi$ are the cylindrical coordinates, $k = 2\pi/\lambda$ and $\lambda$ are the wavenumber and wavelength in the host medium, $w_0$ is the beam parameter and represents the spatial extent of the beam at $z = 0$ that is the beam's minimum-waist plane, and the Rayleigh range is defined as $z_R = \pi w_0^2 / \lambda$. Note that at any given $z$, the azimuthal electric field is maximum at the radius $\rho = \rho_M$ where $\rho_M = w/\sqrt{2}$ [31], and its maximum value is equal to

$$\left| E_\varphi^{\text{APB}}(\rho_M, z) \right| = \left| V \right| \frac{2\rho}{\sqrt{\pi w^2}} e^{-\left(\frac{\rho}{w}\right)^2} \Bigg|_{\rho=\rho_M} = |V| \frac{1}{w}\sqrt{\frac{2}{\pi e}}. \quad (5)$$

The magnetic field of the APB is then calculated via $\mathbf{H} = \nabla \times \mathbf{E}/(i\omega\mu)$ with the electric field of the APB given in Eq. (3). The $z$ component of the magnetic field is

$$H_z^{\text{APB}} = \frac{V}{\eta}\frac{-2i}{\sqrt{\pi^3}}\frac{\lambda}{w^2} e^{-\left(\frac{\rho}{w}\right)^2} \zeta \, e^{-2i\tan^{-1}\left(\frac{z}{z_R}\right)} e^{ikz}\left(1 - \frac{\rho^2}{w^2}\zeta\right), \quad (6)$$

which is not only nonzero on the beam axis ($\rho = 0$) but also reaches its maximum magnitude

$$\left| H_z^{\text{APB}}(\rho = 0) \right| = \frac{|V|}{\eta}\frac{2}{\sqrt{\pi^3}}\frac{\lambda}{w^2}. \quad (7)$$

Here note that the maximum electric field magnitude is inversely proportional to $w$ whereas the maximum longitudinal magnetic field is inversely proportional to $w^2$. Therefore when keeping the power in the APB constant, one can boost maximum magnetic field relatively more than the maximum electric field by focusing the beam to tighter spots. Note that APB excitation ideally provides a $F_Y \to \infty$ on the beam axis, hence it is ideal for obtaining regions with large local field admittance. The radial magnetic field is accordingly found as

$$H_\rho^{\text{APB}} = -\frac{1}{\eta} E_\varphi^{\text{APB}}\left[1 + \frac{1}{kz_R}\frac{\rho^2 - 2w_0^2}{w^2}\right]. \quad (8)$$

It is observed that when $kz_R \gg 1$ the radially polarized magnetic field component follows mainly the same intensity profile as the azimuthally polarized electric field. For very tight beams, however, the second term inside the brackets in (8) becomes non negligible, and a slight difference between the intensity profiles of $H_\rho^{\text{APB}}$ and $E_\varphi^{\text{APB}}$ starts to appear. The formula of the power of the APB, $P$, as a function of $V$ and $w_0/\lambda$ is given in Appendix.

Next we emphasize the relative increase of the longitudinal magnetic field of an APB and its resolution as $w_0$ decreases. In Fig. 2 all the field components of APBs with $w_0 = \lambda$ and $w_0 = 0.5\lambda$ are reported, keeping the power carried in the beams constant and equal to 1 mW. We observe that the longitudinal magnetic field is boosted relatively more than the transverse magnetic and electric field components when the beam is tighter, i.e., when $w_0$ decreases. Therefore just using an APB with a tighter spot is on its own an intriguing way to boost the magnetic field, the magnetic field resolution and the figure of merit $F_Y$. Moreover in the following we will analytically prove that a magnetically polarizable cluster further enhances the magnetic field at its center, almost independently of the beam parameter of the incident APB.

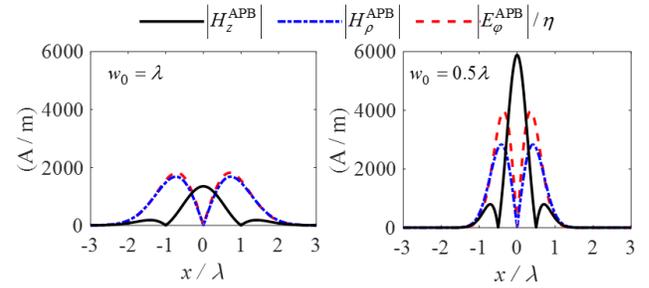

Fig. 2. Field profiles of an APB in vacuum with two different beam parameters, $w_0 = \lambda$ and $w_0 = 0.5\lambda$ at $\lambda = 632$ nm keeping the power in the beam constant and equal to 1 mW (i.e., with $V = 0.972$ V for $w_0 = \lambda$ and $V = 0.891$ V for $w_0 = 0.5\lambda$). Longitudinal magnetic field $H_z^{\text{APB}}$ is boosted in tight beams (i.e., with small beam parameter $w_0$).

We define a figure of longitudinal magnetic field at the minimum waist (i.e., at $z = 0$), which is equal to the longitudinal magnetic field of an APB normalized by the magnetic field of a plane wave whose power intensity is equal to the power of the APB divided by an area $\lambda^2$

$$h_z = |H_z| / \sqrt{\frac{2P}{\eta\lambda^2}}. \quad (9)$$

where $P$ is the power in the APB which possesses the longitudinal magnetic field $H_z$. The figure $h_z$, evaluated at the minimum waist is only a function of $w_0/\lambda$ and its explicit formula is provided in Appendix. In Fig. 3, we report such figure of magnetic field and it clearly shows that the magnetic field of an APB is significantly boosted as the beam parameter decreases, especially when $w_0 < \lambda$.

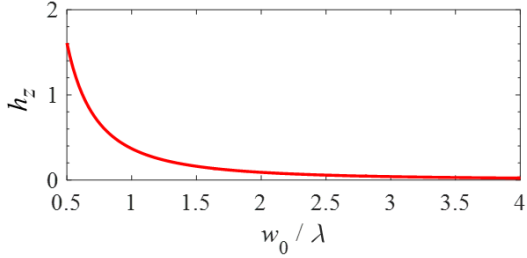

Fig. 3. The nondimensional figure of longitudinal magnetic field defined in (9).

In the following, we introduce the analytical equations used in the single dipole approximation (SDA) calculations for quantifying the two main figures of merit in (1), the magnetic field enhancement and the normalized absolute local field admittance, and then explain the physics behind the capability of the cluster for boosting these two figures of merit. Note that in our characterization of the cluster under APB illumination, we put special emphasis on the magnetic field enhancement, i.e., on $F_H$, achieved in the presence of the ring cluster as a way of boosting natural magnetism, rather than the cluster's magnetic polarizability as done in several previous publications [15].

### III. CLUSTER ANALYTICAL MODEL

We briefly summarize the analytical model utilized in solving a scattering problem where a cluster made of electrically polarizable *plasmonic* nanospheres in homogeneous free space, vacuum, is excited by an external field. For comparison we also consider external field such as a single plane wave and a superposition of two plane waves.

The nano cluster is made of $N$ plasmonic particles at positions $\mathbf{r}_n$, displaced in the $x$-$y$ plane (Fig. 1). For simplicity we assume that each particle is a nanosphere modeled by a scalar (isotropic) electrical polarizability $\alpha_n$ where $n \in \{1,...,N\}$. The local electric field exciting the $n^{\text{th}}$ particle is evaluated as

$$\mathbf{E}^{\text{loc}}(\mathbf{r}_n) = \mathbf{E}^{\text{ext}}(\mathbf{r}_n) + \sum_{\substack{m=1 \\ m \neq n}}^{N} \underline{\mathbf{G}}_{Ep}(\mathbf{r}_n, \mathbf{r}_m) \cdot \mathbf{p}_m , \quad (10)$$

where the superscript 'ext' denotes the external excitation (i.e., the illuminating APB) and $\underline{\mathbf{G}}_{Ep}(\mathbf{r}_n, \mathbf{r}_m)$ is the dyadic Green's function that gives the electric field at $\mathbf{r}_n$ generated by the electric dipole $\mathbf{p}_m$ at $\mathbf{r}_m$, and here we use the fully dynamic and exact expression in Chapter 8 of [33]. Each nanosphere's electric dipole moment is evaluated via the Mie polarizability of the nanosphere and the local field at the nanosphere's center as $\mathbf{p}_n = \alpha_n \mathbf{E}^{\text{loc}}(\mathbf{r}_n)$, where the local field is the sum of the external field and the field scattered by all the other nanospheres. Thus (10) is rewritten in terms of the unknown electric dipole moments as

$$\mathbf{p}_n - \alpha_n \sum_{\substack{m=1 \\ m \neq n}}^{N} \underline{\mathbf{G}}_{Ep}(\mathbf{r}_n, \mathbf{r}_m) \cdot \mathbf{p}_m = \alpha_n \mathbf{E}^{\text{ext}}(\mathbf{r}_n) . \quad (11)$$

By writing (11) for $n = 1, ..., N$, we construct a linear system of $N$ equations as

$$[A] \begin{bmatrix} \mathbf{p}_1 \\ \vdots \\ \mathbf{p}_N \end{bmatrix} = \begin{bmatrix} \alpha_1 \mathbf{E}^{\text{ext}}(\mathbf{r}_1) \\ \vdots \\ \alpha_N \mathbf{E}^{\text{ext}}(\mathbf{r}_N) \end{bmatrix}, \quad (12)$$

where $[A]$ is a $3N \times 3N$ matrix made of $3 \times 3$ sub-blocks $\underline{\mathbf{A}}_{nm}$ with $n,m \in \{1,...,N\}$ that are given by

$$\underline{\mathbf{A}}_{nm} = \begin{cases} \underline{\mathbf{I}} & \text{when } m = n \\ -\alpha_n \underline{\mathbf{G}}_{Ep}(\mathbf{r}_n, \mathbf{r}_m) & \text{otherwise} \end{cases}, \quad (13)$$

where $\underline{\mathbf{I}}$ is the $3 \times 3$ identity matrix. The system of linear equations in (12) is solved for the electric dipole moments under external excitation. Subsequently the electric and magnetic fields scattered by the cluster are evaluated using the fully dynamic and exact dyadic Green's functions $\underline{\mathbf{G}}_{Hp}$ (Section 2.3 in [34]) and $\underline{\mathbf{G}}_{Ep}$ which provide the magnetic and electric fields due to an electric dipole, respectively. The total electric and magnetic fields at an observation point $\mathbf{r}$ are evaluated as

$$\mathbf{E}^{\text{tot}}(\mathbf{r}) = \mathbf{E}^{\text{ext}}(\mathbf{r}) + \sum_{n=1}^{N} \underline{\mathbf{G}}_{Ep}(\mathbf{r}, \mathbf{r}_n) \cdot \mathbf{p}_n,$$
$$\mathbf{H}^{\text{tot}}(\mathbf{r}) = \mathbf{H}^{\text{ext}}(\mathbf{r}) + \sum_{n=1}^{N} \underline{\mathbf{G}}_{Hp}(\mathbf{r}, \mathbf{r}_n) \cdot \mathbf{p}_n. \quad (14)$$

For simplicity, in the following section we assume that all nanospheres are identical and we drop the subscript $n$ in the polarizability symbol.

### IV. PHYSICS OF CLUSTER AZIMUTHAL EXCITATION, RESONANCE AND FIELD ENHANCEMENT

A cluster made of plasmonic nanospheres offers a large degree of flexibility in tuning the magnetic resonance wavelength. In general, we define the *real* magnetic resonance wavelength as the wavelength at which the magnitude of the magnetic dipole moment of the cluster under a time harmonic field peaks. The magnetic dipole moment of the cluster as in Fig. 1 is defined as

$$\mathbf{m} = \frac{i\omega}{2} \sum_{n=1}^{N} \mathbf{r}_n \times \mathbf{p}_n . \quad (15)$$

The cluster has the significant cross-sectional area only on the plane normal to the $z$ axis, i.e. the main magnetic moment of the cluster will be aligned along $z$ under various types of excitations. The magnetic moment of the cluster is proportional to the local magnetic field (assumed not varying significantly over the cluster area). Therefore we define the magnetic polarizability of the cluster as a way of modeling its magnetic response. The magnetic polarizability is in general represented as a tensor, however here we are interested in the $z$-directed magnetic moment induced by the $z$-component of the external magnetic field (i.e., an APB with strong longitudinal magnetic field along its propagation axis $z$). Accordingly the magnetic polarizability of the cluster centered at the origin is

$$\alpha_{zz}^{mm} = \frac{m_z}{H_z^{\text{ext}}(\mathbf{r}=\mathbf{0})}. \quad (16)$$

Due to symmetry, the magnetic dipole moment is generated by circulating electric dipolar moments (Fig. 1), which are excited by an APB. The cluster in Fig. 1 is made of nanospheres equally spaced on a perfect circle with the cluster radius $a = (2r+g)/[2\sin(\pi/N)]$ where $r$ is the nanosphere radius and $g$ is the inter-nanosphere gap distance. The cluster is excited by an ideal APB whose electric field $E_\varphi^{\text{APB}}$ is purely azimuthal, i.e., transverse to $z$ and along the $\varphi$ direction. Under rotational symmetry, all the induced electric dipole moments are polarized azimuthally and equal in magnitude given by

$$p_\varphi = \frac{\alpha E_\varphi^{\text{APB}}(\rho = a)}{1 - \alpha \sum_{n=2}^{N} \hat{\boldsymbol{\varphi}}_1 \cdot \underline{\mathbf{G}}_{Ep}(\mathbf{r}_0, \mathbf{r}_n) \cdot \hat{\boldsymbol{\varphi}}_n}, \quad (17)$$

where $\hat{\boldsymbol{\varphi}}_n = -\sin[(n-1)\delta]\hat{\mathbf{x}} + \cos[(n-1)\delta]\hat{\mathbf{y}}$ with $\delta = 2\pi/N$. In the following the denominator in (17) is called

$$D = 1 - \alpha \sum_{n=2}^{N} \hat{\boldsymbol{\varphi}}_1 \cdot \underline{\mathbf{G}}_{Ep}(\mathbf{r}_1, \mathbf{r}_n) \cdot \hat{\boldsymbol{\varphi}}_n, \quad (18)$$

and it reaches its minimum absolute value at the "magnetic" resonance.

Since the electric dipole moments and the position vectors ($\mathbf{p}_n$ and $\mathbf{r}_n$) lie on the same plane (orthogonal to $z$) the magnetic dipole moment of the cluster would be purely in the $z$ direction based on (15). This leads to an equivalent cluster magnetic dipole moment and the magnetic cluster polarizability as

$$m_z = N\frac{i\omega}{2}ap_\varphi, \quad \alpha_{zz}^{mm} = N\frac{i\omega}{2}a\frac{p_\varphi}{H_z^{\text{APB}}(\mathbf{0})}. \quad (19)$$

By substituting (17) in (19), the cluster magnetic polarizability (16) is found as

$$\alpha_{zz}^{mm} = N\frac{i\omega}{2}a\frac{\alpha}{D}\frac{E_\varphi^{\text{APB}}(\rho = a, z = 0)}{H_z^{\text{APB}}(\mathbf{0})}, \quad (20)$$

The $E_\varphi^{\text{APB}} / H_z^{\text{APB}}$ field ratio is found by looking at (3) and (6):

$$\frac{E_\varphi^{\text{APB}}(\rho, z)}{H_z^{\text{APB}}(\rho = 0, z)} = \eta\frac{\pi\sqrt{2}}{2}\frac{\rho}{\lambda}e^{-\left(\frac{\rho}{w}\right)^2 \xi},$$

$$\frac{E_\varphi^{\text{APB}}(\rho = a, z = 0)}{H_z^{\text{APB}}(\mathbf{0})} = \eta\frac{\pi\sqrt{2}}{2}\frac{a}{\lambda}e^{-\left(\frac{a}{w_0}\right)^2}, \quad (21)$$

and reaches maximum when $a = \rho_M = w_0/\sqrt{2}$.

From (1) the longitudinal magnetic field enhancement at the cluster center is

$$F_H(\mathbf{0}) = \left|\frac{H_z^{\text{ext}}(\mathbf{0}) + H_z^{\text{scat}}(\mathbf{0})}{H_z^{\text{ext}}(\mathbf{0})}\right| = \left|1 + \frac{H_z^{\text{scat}}(\mathbf{0})}{H_z^{\text{ext}}(\mathbf{0})}\right|, \quad (22)$$

where $H_z^{\text{ext}} \equiv H_z^{\text{APB}}$. The scattered magnetic field enhancement is

$$\frac{H_z^{\text{scat}}(\mathbf{0})}{H_z^{\text{APB}}(\mathbf{0})} = N\frac{\omega k}{4\pi}\frac{e^{ika}}{a}\left[1 + \frac{i}{ka}\right]\frac{p_\varphi}{H_z^{\text{APB}}(\mathbf{0})}$$
$$= N\frac{\omega k}{4\pi}\frac{e^{ika}}{a}\left[1 + \frac{i}{ka}\right]\frac{\alpha}{D}\frac{E_\varphi^{\text{APB}}(\rho = a, z = 0)}{H_z^{\text{APB}}(\mathbf{0})}, \quad (23)$$

which is the most significant term in the magnetic field enhancement reported in (22) when maximized. In the expressions in (20) and (23) we see the common term $E_\varphi^{\text{APB}}/H_z^{\text{APB}}(\mathbf{0})$ given in (21).

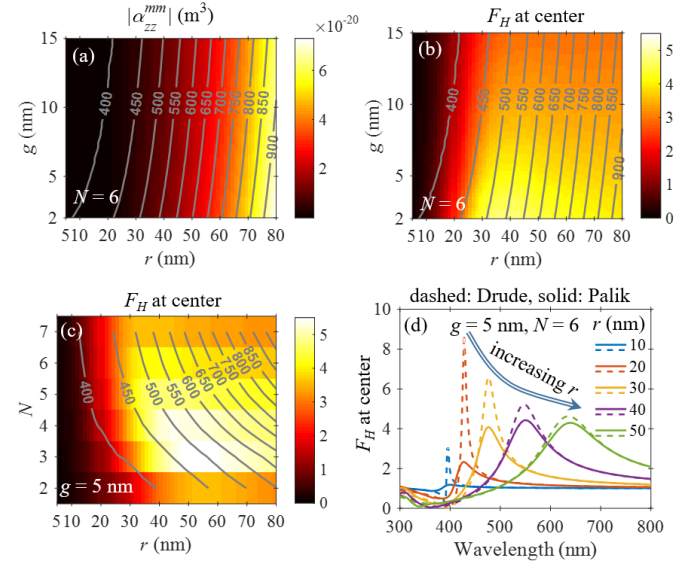

Fig. 4. (a) Cluster magnetic polarizability and (b) magnetic field enhancement $F_H$ at the cluster's center (purely $z$-polarized owing to the symmetry of the cluster geometry and APB excitation), at the respective wavelengths where each peaks, versus nanosphere radius $r$ and gap $g$ between neighboring nanospheres when $N = 6$. (c) $F_H$ at the wavelength where it peaks, versus $N$ and $r$ when gap is fixed at $g = 5$ nm. The superimposed iso-wavelength contours annotated with the wavelengths in nm denote the wavelength at which the reported quantity peaks. (d) $F_H$ versus wavelength for various $r$ and the effect

of losses on the maximum magnetic field enhancement at the wavelengths where it peaks using two different silver permittivity functions, Drude's model and experimental Palik data.

The optimum cluster design that maximizes either one of the magnetic polarizability and the magnetic field enhancement does not necessarily maximize the other one. In this paper the main quantity of interest is not in the magnetic moment or the magnetic polarizability of the cluster but rather the magnetic field enhancement in the central area of the cluster. The magnetic field enhancement at the cluster center is purely due to enhancement of the longitudinal (z-directed) magnetic field and we will use the figure of merit ($F_H$) to quantify it.

We plot in Fig. 4(a, b) the peak of the cluster magnetic polarizability $\alpha_{zz}^{mm}$ and the peak of the magnetic field enhancement $F_H$ of a cluster made of 6 silver nanospheres excited by an APB with beam parameter $w_0 = \lambda$, versus nanosphere radius $r$ and inter-sphere gap $g$ using SDA model. The cluster is placed at the APB minimum waist plane (i.e, at $z = z_0$) and nanospheres' silver is described by the "Palik" permittivity function taken from [35]. The iso-wavelength contours, annotating the wavelength in nm at which these peaks occur is shown in Fig. 4(a-c). The colormaps are generated by calculating the peak values (that are wavelength dependent) for each pair of $g$, $r$ or $N$, $r$ parameters. Note that as the nanospheres radius increases the peak magnetic polarizability increases monotonically *in the reported range*, whereas the field enhancement peaks at a certain nanosphere radius, for each gap distance. Moreover the magnetic polarizability and magnetic field enhancement peaks occur almost at the same wavelength. As the gap $g$ increases and $r$ is kept constant, the peak magnetic polarizability and the peak magnetic field enhancement decreases. Therefore small gaps are important for achieving strong resonances.

In Fig. 4(c) the magnetic field enhancement peak is plotted versus $N$, the number of nanospheres, and nanospheres radius $r$ where the inter-sphere gap is fixed at $g = 5$ nm. Similarly there appears to be an optimum nanosphere radius for each $N$. Also it is observed that $N = 3$ or $4$ lead to largest field enhancement values, however it may be preferable to use $N = 6$ which leads to a larger accessible area of magnetic field enhancement inside the cluster.

It is observed in Fig. 4(a,b) that for a specified $N$ and when keeping $r$ constant, the resonance wavelength decreases very *slightly* as $g$ increases. On the other hand, when $N$ and $g$ are kept constant, the resonance wavelength of the cluster increases significantly as $r$ increases. In Fig. 4(c) where $g$ is fixed to 5 nm, the resonance wavelength increases notably, either with increasing $r$ and keeping $N$ constant or with increasing $N$ and keeping $r$ constant. Recalling that the cluster radius is $a = (2r + g)/[2\sin(\pi/N)]$, as either $r$ or $N$ gets larger the cluster radius increases. Furthermore when $r \gg g$, $a$ becomes proportional to $r$ as $a \approx r/\sin(\pi/N)$. Therefore the *main* trend in Fig. 4(a-c) is that the resonance wavelength increases when the cluster radius $a$ increases due to an increase of $r$ or $N$.

Finally in Fig. 4(d) the magnetic field enhancement versus wavelength is plotted for various nanosphere radii using the Palik permittivity function [35] (as in the other map plots in Fig. 4) and Drude's model [36] which underestimate losses in silver at smaller wavelength range and leads to larger magnetic field enhancement values.

The optimum magnetic field enhancement occurs at certain nanosphere radius and number of nanospheres $N$ whereas the magnetic polarizability is monotonically increasing with nanosphere radius in the reported range. Therefore it is apparent that different design considerations apply to maximize either magnetic field enhancement or magnetic polarizability. The cluster is equivalent to a circulating electric current as an effective magnetic dipole. Therefore the radius $a$ of the cluster is a crucial parameter because a larger cross-sectional area of the cluster leads to a larger magnetic polarizability but not necessarily to a larger magnetic field enhancement. Next we interpret the observations based on the analytical formulas.

The term $D$ at the denominator in (18) determines the resonance of the cluster, and it appears in both the magnetic polarizability (20) and the scattered magnetic field enhancement (23). These quantities are proportional to the electric polarizability $\alpha$ of each nanosphere which grows as $\alpha \propto r^3$. Keep in mind that as $r$ increases, the cluster radius $a = (2r + g)/[2\sin(\pi/N)]$ also increases. Next, the term $E_\varphi^{\mathrm{APB}}(a)/H_z^{\mathrm{APB}}(0)$ in (21) is common for both the magnetic polarizability in (20) and the scattered magnetic field enhancement in (23). Here we emphasize that, when neglecting the signature of $D$, both $\alpha_{zz}^{mm}$ and the scattered field enhancement in (20) and (23) tend to grow with the cluster radius $a$ (i.e., with increasing $r$ or $N$), assuming that $a^2 \ll w_0^2$. When this assumption is not verified, the exponential function $\exp[-(a/w_0)^2]$ in the common term (21) limits the growth of $E_\varphi^{\mathrm{APB}}/H_z^{\mathrm{APB}}$. We now focus on the cluster radius that maximizes either $\alpha_{zz}^{mm}$ or the scattered field enhancement $H_z^{\mathrm{scat}}(0)/H_z^{\mathrm{APB}}(0)$. To do so we utilize the dimensionless ratio

$$\frac{\left|H_z^{\mathrm{scat}}(0)/H_z^{\mathrm{APB}}(0)\right|}{\left|\alpha_{zz}^{mm}\right|/\lambda^3} = \frac{4\pi^2}{(ka)^2}\sqrt{1 + \frac{1}{(ka)^2}} \qquad (24)$$

in order to assess the relative dependence of $\alpha_{zz}^{mm}$ and the scattered field enhancement on the cluster radius $a$. The ratio in (24) does not depend on the term $D$ in (18) and it grows when $a$ decreases. It is clear that among different resonant designs, the cluster with smaller cluster radius tends to have relatively large field enhancement compared to the absolute magnetic polarizability normalized by $\lambda^3$. As $a$ increases, the magnetic cluster polarizability grows faster than the magnetic field enhancement with an extra $a^3$ factor dependence when $(ka)^2 \ll 1$. On the other hand as $a$ increases further, the term $\exp(-a^2/w_0^2)$ causes both quantities in (20) and (23) to reach a peak and then decrease. However $\alpha_{zz}^{mm}$ in (20) and the scattered field enhancement in (23) reach maximum at different cluster radii. We observe in Fig. 4(a,b) that when keeping $g$ and $N$ constant, the magnetic field enhancement $F_H$ reaches peak value for certain $r$ values in the reported range whereas the magnetic polarizability $\alpha_{zz}^{mm}$ grows monotonically in the reported $r$ range and it is expected to reach peak value at a large $r$ out of the reported range. Thus, $\alpha_{zz}^{mm}$ peaks at a larger cluster radii than the one where $F_H$ peaks, when $g$ and $N$ are kept constant.

The cluster here is seen as a current loop and this analogy helps us conceive the physics behind maximizing the magnetic field enhancement rather than the magnetic polarizability. The magnetic polarizability of a current loop is proportional to the loop area squared, thus proportional to $a^4$. Moreover the current induced on a loop is proportional to the area and the incident magnetic field. Accordingly the magnetic field at the center of a current loop is proportional to the loop current but inversely proportional to $a$, thus the magnetic field enhancement is proportional to the loop radius $a$. Eventually we observe a factor difference of $a^3$ in the dependences of the magnetic polarizability and the magnetic field enhancement on $a$ which is in agreement with the formula in (24) for a cluster when $(ka)^2 \ll 1$.

Having discussed the characterization of the cluster magnetic resonance and the magnetic field enhancement, in the following we stress the advantages of exciting the nano cluster with an APB compared to other possible excitation schemes. In Fig. 5, we compare the two figures of merit, the magnetic field enhancement $F_H$ and the normalized absolute local field admittance $F_Y$, both evaluated at the cluster plane using three different excitation schemes: (i) TE (with respect to $z$) plane wave propagating in the $x$ direction, (ii) two antisymmetric plane waves propagating in +/- $x$ directions, (iii) APB with $w_0 = \lambda$ whose beam axis coincides with the cluster axis (the $z$ axis). The case with two plane waves have vanishing electric field at the cluster center. All the excitation schemes excite the magnetic resonance significantly and lead to a magnetic field enhancement around 4.2. Furthermore, the single plane wave case does not lead to an increased local admittance, the two plane-waves provide high local field admittance, and the APB excitation leads to the largest value and widest area of enhanced local admittance, representing a wide magnetic-dominant region. Note that even though the two antisymmetric plane-wave scheme also results in a large local field admittance in the cluster's center, it is difficult to phase synchronize these plane waves in practical cases such that their electric fields cancel out exactly at the cluster's center, whereas vanishing electric field at the cluster center is a natural property of the APB.

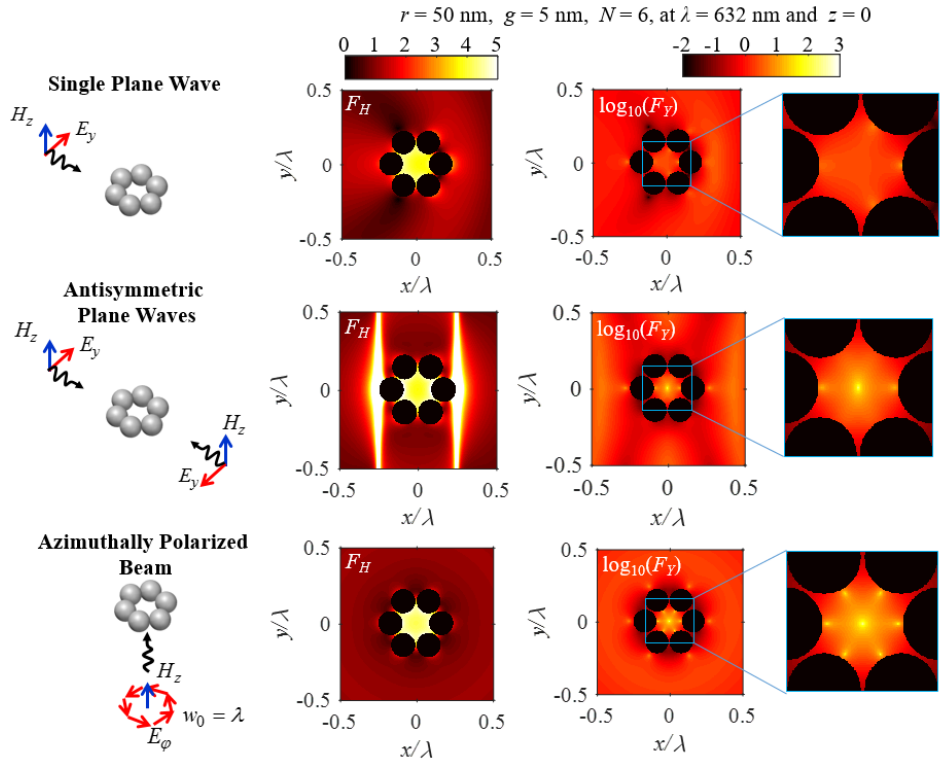

Fig. 5. Local magnetic field enhancement $F_H$ (First column of plots) and normalized local field admittance $F_Y$ (second column of plots), both evaluated on the transverse symmetry plane of the cluster for three different excitations: (i) single plane-wave incidence, (ii) two antisymmetric plane-wave incidence, (iii) normally incident APB.

## V. FIELD CHARACTERISTICS VERSUS APB EXCITATION

At this point we have only plotted the characteristic field maps of the nano cluster excited by an APB with $w_0 = \lambda$. In this section, we first characterize the cluster magnetic response versus the beam parameter $w_0$ of the APB and then report the figures of merit at several planes, when illuminated by an APB propagating in $+z$ direction as in Fig. 6.

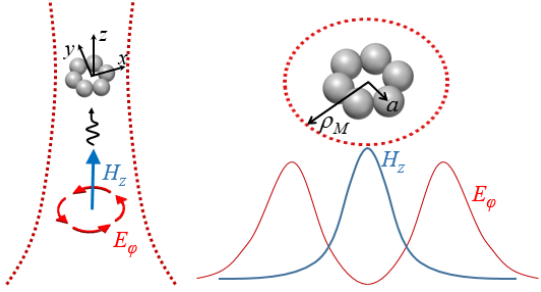

Fig. 6. Illustration of the cluster located at the minimum waist plane of an APB propagating in the $+z$ direction. The radial distance $\rho_M$ where the electric field is maximum is denoted by a dashed line.

In general, the figures of merit investigated quantitatively in Sec. IV depend on the beam parameter $w_0$ of the illuminating APB that determines also the radial location $\rho_M = w_0 / \sqrt{2}$ of the maximum of the electric field. Note that the cluster considered in this section with parameters $r = 50$ nm, $g = 5$ nm, $N = 6$ resonates at a wavelength of $\lambda = 632$ nm, and its radius is $a \approx \lambda/6$. In order to ensure that the APB electric field maximum coincides with the cluster radius, so $\rho_M = a$, the beam parameter of the APB should be $w_0 \approx 0.24\lambda$. However the field features of such an APB is beyond the diffraction limit, and cannot constitute a propagating beam as investigated in [31]. However, it is still important to assess the impact of the beam parameter $w_0$, i.e. the spatial extent and amplitude distribution of the excitation field, on the magnetic field enhancement of the cluster-APB system. Therefore we report in Fig. 7 the magnetic cluster polarizability $\alpha_{zz}^{mm}$ and the magnetic field enhancement $F_H$ at the cluster center versus beam parameter. In both plots we observe the signature of $E_\varphi^{\text{APB}} / H_z^{\text{APB}}(\mathbf{0})$ term [appearing as a function of $w_0$ in (20) and (23)] as a slight decrease with $w_0$ since the exponential term in (24) becomes significant. It is shown that for $w_0 > \lambda$ both quantities plotted in Fig. 7 saturate, and around $w_0 \approx \lambda$ the reported quantities take values close to the saturated ones. Note here that even though the APB's maximum electric field location, as illustrated in Fig. 6, moves farther from the cluster radius, when $w_0$ is increased; the magnetic field enhancement does not change significantly. Smaller $w_0$ indicates tighter field features that start to be comparable to the cluster size, thus the magnetic polarizability and the magnetic field enhancement decrease slightly as $w_0$ decreases. However the slight decrease in the magnetic field enhancement $F_H$ with decreasing $w_0$ does not mean that the APBs with tighter features should be avoided. In fact, we recall that $F_H$ is defined in (1) as the ratio of the total field over the incident (external) field, and despite the slight decrease of $F_H$ with decreasing $w_0$, the incident (external) magnetic field of a tighter beam is much stronger, assuming that the power of the beam is kept constant. This is easily understood by looking at the magnetic field of the incident APB in Fig. 2 for a tightly focused APB ($w_0 = \lambda/2$) and a weakly focused APB ($w_0 = \lambda$). It is clear from Fig. 2 that the incident magnetic field with $w_0 = \lambda/2$ is almost 3 times the one with $w_0 = \lambda$, whereas in Fig. 7 we observe only a 10% drop in enhancement from $w_0 = \lambda$ to $w_0 = \lambda/2$. Eventually, we still stress that tighter beams lead to larger total magnetic fields. In Fig. 7, the sweep of $w_0$ is started at $w_0 = \lambda/2$ because as the beam parameter $w_0$ decreases to values smaller than $\lambda/2$ the plane wave spectrum of its field starts to extend over to the evanescent spectrum and these beams are no more composed of a spectrum of propagating waves and the paraxial approximation in (1) loses accuracy [31]. The field features throughout this paper are calculated using an APB with $w_0 = \lambda$ which represents a self-standing beam whose field spectrum is only confined to the propagating plane-wave spectrum.

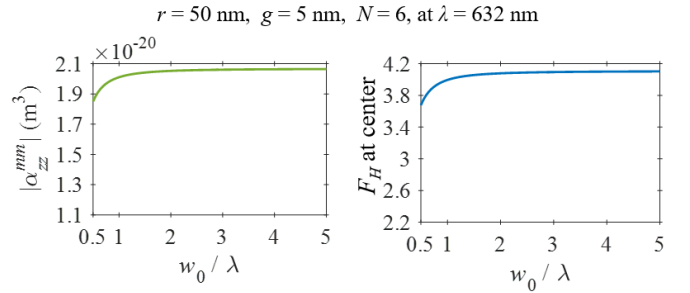

Fig. 7. $\alpha_{zz}^{mm}$ and $F_H$ at the origin, i.e., center of the cluster, versus the beam parameter of APB.

In Fig. 8(a-c), we report $F_H$, $F_E$, and $F_Y$ and then along the $x$ and $y$ axes at several $z$-planes, from $z = -0.5\lambda$ to $z = 0.5\lambda$ where we assume the cluster is centered at the minimum waist plane $z = 0$ and the APB is incident from below as in Fig. 6. It is observed that the field enhancement features are mainly confined to the cluster plane, and the normalized absolute local field admittance $F_Y$ is maximum around the $z$ axis and in contrast to the other figures of merit, it maintains its large value at different $z$ values. It is important to note here that $F_Y$ ideally tends to infinity on the $z$ axis, and is there truncated (for graphic representation) at a maximum of $10^3$ (or 60 dB) in the plots. Lastly we report the magnetic field enhancement $F_H$ along the $z$ axis, whose maximum value occurs at $z = 0$. Importantly, we see the destructive interference signature of the incident magnetic field and the scattered one as a minimum at $z = -0.25\lambda$.

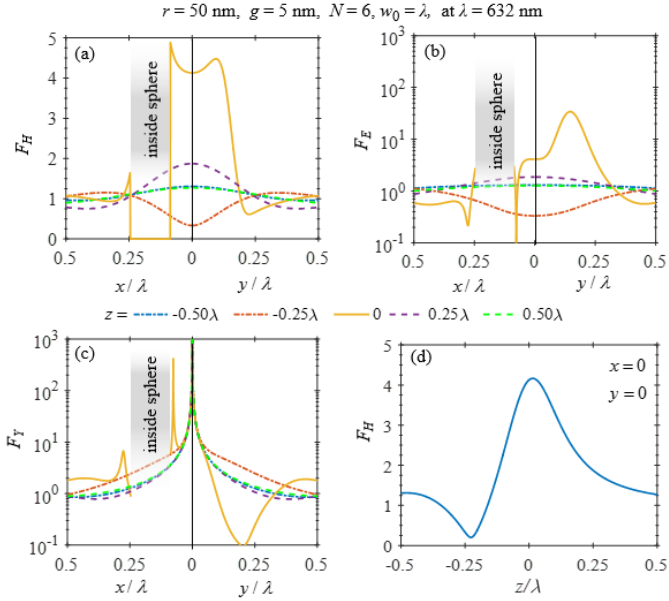

Fig. 8. (a) Magnetic and (b) electric field enhancement ($F_H$ and $F_E$). (c) Normalized absolute local field admittance $F_Y$ versus $x$ and $y$ at various $z$ planes. (d) The magnetic field enhancement on $z$ axis showing destructive interference at $z \approx -0.25\lambda$ and a maximum at $z/\lambda \approx 0$.

## VI. EFFECT OF CLUSTER DEFECTS AND BEAM ALIGNMENT ON FIGURES OF MERIT

In the previous section, we have shown that the normalized absolute local field admittance $F_Y$ around the center of the nano cluster excited by the APB is ideally very large alongside a magnetic field enhancement $F_H$ around 4.2. In addition to an increase of $F_H$ due to the nano cluster, stronger incident magnetic field is also achieved with tighter APBs. However it is supposable that the perfect alignment of the beam axis with the cluster axis and also the ideal symmetry of the circular cluster may not be easily achieved in practical applications. In this section, we provide a short assessment of the sensitivity of advantages of the APB illumination obtained with the proposed setup in Fig. 1 with respect to some defect scenarios.

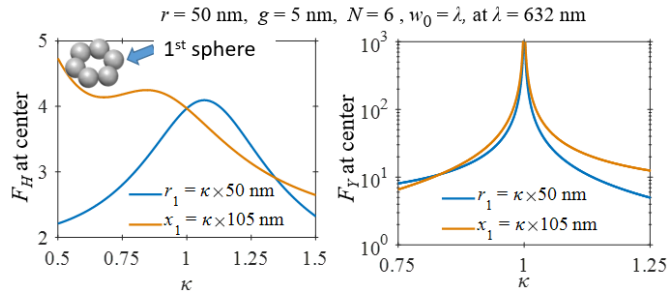

Fig. 9. $F_H$ and $F_Y$ at the origin, i.e., center of the cluster, versus two defect scenarios, either (i) only the radius of 1st sphere in the cluster, or (ii) only the position of the 1st sphere in the cluster is scaled by a factor $\kappa$ with respect to the ideally symmetric cluster. The reference nanosphere radius is set to 50 nm and the reference distance of the nanospheres from the origin is equal to 105 nm.

We investigate two possible defect scenarios of nano clusters where the nanosphere on the +x axis (1st sphere) (i) has a different radius than the rest of the nanospheres, and (ii) is displaced along the x axis. To examine the effect of these defects in the nano cluster on its figures of merits, the radius and position of the 1st nanosphere in the regular symmetric cluster are, respectively, scaled by a coefficient $\kappa$, namely equal to $r_1 = \kappa(50\,\text{nm})$ and $x_1 = \kappa(105\,\text{nm})$, respectively. It is observed in Fig. 9 that by scaling the 1st nanosphere's radius with $\kappa = 0.9$ to $\kappa = 1.25$, one still has a magnetic field enhancement $F_H$ at the cluster center larger than 90% of its nominal value with $\kappa = 1$. In addition, the magnetic field enhancement $F_H$ increases as the nanosphere is placed closer the cluster center. In contrast to the magnetic field enhancement, the magnetic to electric field ratio $F_Y$ at the cluster center shows a very strong dependence on the physical defects in the cluster. We recall that in an ideal symmetric setting $F_Y = \infty$ when $\kappa = 1$. However even 10% variation in the radius or the position of the 1st sphere can lead to a decrease to $F_Y = 15$ due to the loss of the radial symmetry in the cluster-APB setup. For variations within $0.83 < \kappa < 1.13$ one still has $F_Y > 10$.

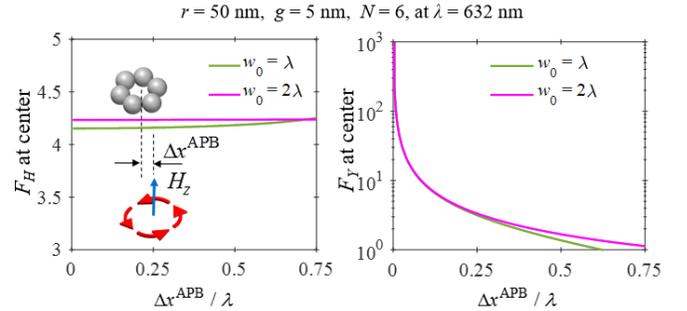

Fig. 10. Effect of beam axis displacement from the center of the cluster $\Delta x^{\text{APB}}$ on the magnetic field enhancement $F_H$ (left) and the normalized absolute local field admittance $F_Y$ (right) at the cluster center. It is shown that independently of the beam parameter $w_0$ the field enhancement is resilient to the beam alignment whereas the normalized absolute local field admittance is highly sensitive, and remains larger than 10 at the cluster center for $0 \leq \Delta x^{\text{APB}} < 0.08\lambda$.

Next, we plot in Fig. 10, the magnetic field enhancement $F_H$ and normalized magnetic to electric field ratio $F_Y$ at the cluster center versus the *displacement* $\Delta x^{\text{APB}}$ of the APB beam axis along the +x direction from the center of the circular nano cluster. While the magnetic field enhancement $F_H$ is not strongly affected by the offset of the beam axis, the normalized local admittance $F_Y$ drops significantly. However for small displacements within $0 \leq \Delta x^{\text{ABP}}/\lambda < 0.08$ one still has $F_Y > 10$ at the cluster center, since ideally one has $F_Y = \infty$ there. Recalling that the spatial extent of the APB depends on the beam parameter $w_0$, in Fig. 10 we report the figures of merit for two different values $w_0$. We stress that different choices of the beam parameter do not create a significant difference on the figures of merit when it comes to the effects of beam's misalignment $\Delta x^{\text{APB}}$.

## VII. Conclusion

A circular cluster of nanoparticles excited by an azimuthally polarized beam (APB) is utilized as a magnetic nanoprobe for enhancing the magnetic near-field and the spatial resolution of the enhanced magnetic field in a magnetic-dominant region. In the same region a huge local field admittance is achieved, much larger than that of a plane wave, meaning that the magnetic to electric field ratio is very high. We demonstrate that large magnetic field enhancement is robust to small physical defects in the nano cluster and to small misalignments of the APB with respect to the cluster's center, though the latter decreases the local field admittance. In this paper circular clusters of nanospheres as magnetic nanoprobes excited by APBs have been studied as an example, but similar conclusions are expected to hold for other magnetic nanoprobes with symmetry properties. Moreover any required improvement of the model regarding specific fabrication methods and experimental setups (for example the presence of a substrate) should be accounted for in future studies. We remind that different types of magnetic nanoprobes such as silicon spheres or clusters made of different geometries of nanoparticles may provide advantages in experimental setups, tuning wavelength of operation and controlling the magnetic field enhancement level and the area of the magnetic-dominant region. The enhanced magnetic fields in magnetic-dominant regions with resolutions beyond the diffraction limit obtained using magnetic nanoprobes may prove useful in optical spectroscopy and microscopy applications based on detection of magnetic field interacting with matter.

## Appendix

The power $P$ carried by the beam in the $+z$ direction, in the figure of longitudinal magnetic field defined in (9) is given by

$$P = \frac{1}{2} \int_0^{2\pi} \int_0^{\infty} \text{Re}\left\{ -E_\varphi^{\text{APB}} \left( H_\rho^{\text{APB}} \right)^* \right\}_{z=0} \rho \, d\rho \, d\varphi$$

$$= \frac{|V|^2}{2\eta} \left( 1 - \frac{1}{2\pi^2 \left( \frac{w_0}{\lambda} \right)^2} \right) \quad (25)$$

whose derivation is shown in [31]. Therefore $h_z$ in (9), evaluated at $z = 0$ is

$$h_z = \frac{1}{\left(\frac{w_0}{\lambda}\right)^2} \frac{\sqrt{\pi^3}}{\sqrt{1 - \frac{1}{2\pi^2 \left(\frac{w_0}{\lambda}\right)^2}}}, \quad (26)$$

and it is a function of only $w_0 / \lambda$.